\documentclass[lettersize,journal]{IEEEtran}
\usepackage{graphicx}
\usepackage{subfigure}
\usepackage{subeqnarray}
\usepackage{bm}
\usepackage{amsfonts}
\usepackage{amsthm}
\usepackage{amssymb}
\usepackage{makecell}
\usepackage{calc}
\usepackage{color}
\usepackage{mathrsfs}
\usepackage{flushend}
\usepackage{url}
\usepackage{caption}
\usepackage{subfloat}
\usepackage{color}
\usepackage{textcomp}
\usepackage{stfloats}
\usepackage{amsmath,amsfonts}
\usepackage{algorithmic}
\usepackage{algorithm}
\usepackage{array}
\usepackage[caption=false,font=normalsize,labelfont=sf,textfont=sf]{subfig}
\usepackage{textcomp}
\usepackage{stfloats}
\usepackage{url}
\usepackage{verbatim}
\usepackage{graphicx}
\usepackage{cite}
\usepackage[outdir=./]{epstopdf}
\usepackage{booktabs}

\begin{document}

\title{Secure ISAC MIMO Systems:\\ Exploiting Interference With Bayesian\\ Cram{\'e}r-Rao Bound Optimization}

\author{Nanchi Su,~\IEEEmembership{Graduate~Student~Member,~IEEE}, Fan Liu,~\IEEEmembership{Senior~Member,~IEEE}, \\Christos Masouros,~\IEEEmembership{Fellow,~IEEE}, George C. Alexandropoulos,~\IEEEmembership{Senior~Member,~IEEE}, \\Yifeng Xiong,~\IEEEmembership{Member,~IEEE}, and Qinyu Zhang,~\IEEEmembership{Senior~Member,~IEEE}

        % <-this % stops a space
%\thanks{{This work was supported in part by the Engineering and Physical Sciences Research Council (EPSRC) under Grant EP/S028455/1, in part by the National Natural Science Foundation of China under Grant 62101234, Grant U20B2039, Grant 61831008 and Grant 62027802, in part by the Young Elite Scientist Sponsorship Program by CAST under Grant No. YESS20210055, and in part by the China Scholarship Council (CSC). \textit{(Corresponding author: Fan Liu, Qinyu Zhang.)}}}
\thanks{N. Su is with Guangdong Provincial Key Laboratory of Aerospace Communication and Networking Technology, Harbin Institute of Technology (Shenzhen), Shenzhen  518055, China; with the School of System Design and Intelligent Manufacturing, Southern University of Science and Technology, Shenzhen 518055, China; and also with the Department of Electronic and Electrical Engineering, University College London, London WC1E 7JE, U.K. (e-mail: nanchi.su.18@ucl.ac.uk)}
\thanks{F. Liu is with the School of System Design and Intelligent Manufacturing, Southern University of Science and Technology, Shenzhen 518055, China (e-mail: liuf6@sustech.edu.cn).}
\thanks{C. Masouros is with the Department of Electronic and Electrical Engineering, University College London, London WC1E 7JE, U.K. (e-mail: chris.masouros@ieee.org).}
\thanks{G. C. Alexandropoulos is with the Department of Informatics and Telecommunications, National and Kapodistrian University of Athens, 15784 Athens, Greece (e-mail: alexandg@di.uoa.gr).}
\thanks{Y. Xiong is with the School of Information and Electronic Engineering, Beijing University of Posts and Telecommunications, Beijing, 100876, China (e-mail: yifengxiong@bupt.edu.cn).}
\thanks{Q. Zhang is with the Guangdong Provincial Key Laboratory of Aerospace Communication and Networking Technology, Harbin Institute of Technology (Shenzhen), Shenzhen 518055, China, and also with Peng Cheng Laboratory, Shenzhen 518055, China (e-mail: zqy@hit.edu.cn).}
}

% The paper headers
%\markboth{Journal of \LaTeX\ Class Files,~Vol.~14, No.~8, August~2021}%
%{Shell \MakeLowercase{\textit{et al.}}: A Sample Article Using IEEEtran.cls for IEEE Journals}

%\IEEEpubid{0000--0000/00\$00.00~\copyright~2021 IEEE}
% Remember, if you use this you must call \IEEEpubidadjcol in the second
% column for its text to clear the IEEEpubid mark.

\maketitle

\begin{abstract}
In this paper, we present a signaling design for secure integrated sensing and communication (ISAC) systems comprising a dual-functional multi-input multi-output (MIMO) base station (BS) that simultaneously communicates with multiple users while detecting targets present in their vicinity, which are regarded as potential eavesdroppers. In particular, assuming that the distribution of each parameter to be estimated is known \textit{a priori}, we focus on optimizing the targets' sensing performance. To this end, we derive and minimize the Bayesian Cram{\'e}r-Rao bound (BCRB), while ensuring certain communication quality of service (QoS) by exploiting constructive interference (CI). The latter scheme enforces that the received signals at the eavesdropping targets fall into the destructive region of the signal constellation, to deteriorate their decoding probability, thus enhancing the ISAC's system physical-layer security (PLS) capability. To tackle the nonconvexity of the formulated problem, a tailored successive convex approximation method is proposed for its efficient solution. Our extensive numerical results verify the effectiveness of the proposed secure ISAC design showing that the proposed algorithm outperforms block-level precoding techniques.
\end{abstract}


\section{Introduction}\label{sec1}
\IEEEPARstart{F}{uture} radar and communication (R\&C) systems will operate at higher frequencies with larger bandwidth, while possibly exploiting massive antenna arrays and multi-functional reconfigurable intelligent surfaces (RIS), resulting in striking similarities between R\&C systems, including the hardware architecture, channel characteristics, and signal processing methods \cite{liu2023seventy, chepuri2023integrated}. This provides unique opportunities to develop co-design techniques aiming at improving the mutual performance gain of both systems. Meanwhile, with the emergence of smart cities, Internet of Things (IoT) networks, and other advanced applications, the integration of sensing and communication (S\&C) systems is being seen as a transformative technology, enabling autonomous vehicle networks, activity recognition, and unmanned aerial vehicle (UAV) \cite{liu2022survey}. In light of the above, the need for seamless cooperation between S\&C promotes the technical development of integrated sensing and communication (ISAC) systems. 

The utilization of a communal spectrum frequency band, coupled with the intrinsic broadcasting characteristics of wireless transmission, introduces substantial security vulnerabilities in ISAC systems \cite{su2023sensing,su2020secure, alexandropoulos2023counteracting}. In conventional wireless communication systems, security designs are predominantly concerned at the physical layer and the network layer. Compared with network layer security (NLS), physical layer security (PLS) does not require complex cryptographic techniques or key distribution, reducing overhead and complexity. Moreover, PLS may provide a base level of security guarantee even when other layers are compromised, because it leverages the physical characteristics of wireless channels, which are independent of security at other layers of the communication stack. 

The PLS in ISAC systems has been widely studied in recent years. Initially, the artificial noise (AN) is deployed to interfere with eavesdroppers by maximizing the secrecy rate, thus the target/eavesdropper is unable to decode the received signal. To this end, the confidential information is prevented from being intercepted by the target/eavesdropper \cite{su2020secure,li2022improving,ren2022optimal, liu2017securing}. Moreover, the directional modulation (DM) technique, which is based on the principle of constructive interference (CI), has been deployed to design the transmit signal at a symbol level \cite{liu2020secure,wei2020energy,alodeh2015constructive}. In DM, as a step further from AN design, the signals received at multiple eavesdropping targets (Eves) are enforced to fall into the destructive constellation region for further PLS improvements, which leverages destructive interference (DI) as a PLS measure. In particular, the CI-DI technique enables direct alteration of the amplitude and phase of signals at both intended users and potential Eves. Consequently, this paradigm promotes an enhanced symbol error rate (SER) for communication users (CUs), while deteriorating the decoding probability at potential eavesdroppers.

In this work, we consider the estimation task of random parameters of multiple targets, where the prior distribution of parameters is assumed to be known \textit{a priori}. This is common in a number of practical scenarios, such as vehicle tracking, environmental monitoring, etc. Towards that aim, we then evaluate the sensing performance utilizing the lower bound of the unbiased estimation, i.e., Bayesian Cram\'er-Rao Bound (BCRB). Specifically, we formulate a novel signaling design problem that aims to minimize the BCRB, while guaranteeing a predefined quality of service (QoS) at the multiple CUs, by deploying the CI technique and improving the PLS by constraining the received signals at targets/Eves in the destructive constellation region. In the numerical results section, we verify the effectiveness of the proposed algorithm. Moreover, we explore the impact of the \textit{a priori} distribution of the parameters on the radar beampattern as well as the performance tradeoff between the sensing and communication operations. 

\emph{Notations}: Unless otherwise specified, matrices are denoted by bold uppercase letters (i.e., $\mathbf{X}$), vectors are represented by bold lowercase letters (i.e., $\mathbf{x}$), and scalars are denoted by normal font (i.e., $\alpha$). Subscripts indicate the location of the entry in the matrices or vectors (i.e., $s_{i,j}$ and $l_n$ are the $(i,j)$-th and the \emph{n}-th element in $\mathbf{S}$ and $\mathbf{l}$, respectively). $\operatorname{tr}\left(\cdot\right)$ and $\operatorname{vec}\left(\cdot\right)$ denote the trace and the vectorization operations. $\left(\cdot\right)^T$, $\left(\cdot\right)^H$ and $\left(\cdot\right)^*$ stand for transpose, Hermitian transpose and the complex conjugate of the matrices, respectively. $\left\| \cdot\right\|$, $\left\| \cdot\right\|_{\infty}$ and $\left\| \cdot\right\|_F$ denote the $l_2$ norm, infinite norm and the Frobenius norm respectively. $\mathbb{E}\left\{ \cdot  \right\}$ denotes the statistical expectation.

\section{Signal Model}\label{sec2}

We consider a downlink multi-user multi-input single-output (MU-MISO) wireless system, where the dual-functional multi-input multi-output (MIMO) base station (BS) is capable of detecting multi-targets simultaneously with data transmission. The targets are treated as potential Eves of the communication information. The BS is equipped with $N_t$ transmit antennas and $N_r$ receive antennas, enabling communication with $K_{cu}$ single-antenna users and detection of $K_{tar}$ targets of interest{\footnote{From the sensing side, we assume that one sub-array (consisting of $N_t$ antennas) is deployed to transmit signals and another sub-array (comprising $N_r$ antennas) is deployed to receive signals. These sub-arrays are co-located at the BS and operated simultaneously to transmit the dual-function signal and receive its echoes for monostatic sensing. In principle, the transmitted signal will interfere with the reflected echoes (arriving with a round-trip propagation delay) at the receive sub-array, creating a self-interference signal at the BS. This is a typical problem in full-duplex (FD) BSs used for simultaneous communications and sensing \cite{smida2023full}. Fortunately, there exist various approaches for efficiently suppressing self-interference below the noise floor in multi-antenna FD systems, ranging from isolation between the transmit and receive arrays to joint digital and analog beamforming and interference cancellation techniques. Capitalizing on this, in the paper, we neglect the impact of the self-interference assuming that it can be efficiently handled via the state-of-the-art approaches \cite{smida2023full,alexandropoulos2022full}.}}. Below we elaborate on the signal models of both radar and communication systems, respectively.

Let $\mathbf{X} \in \mathbb{C}^{N_t \times L}$ denote the dual-functional signal matrix, where ${\mathbf{X}} = \left[ {{\mathbf{x}}\left[ 1 \right],{\mathbf{x}}\left[ 2 \right], \ldots, {\bf{x}}\left[ L \right]} \right]$, each element of which denotes the transmit signal vector at the $l$-th time slot with $l=1,2,\ldots,L$. Then, the received signal at each $k$-th single-antenna CU, with $k=1,2,\ldots,K_{cu}$, at the $l$-th time slot is given as
\begin{equation}
    {y_{\rm{CU},k}} \left[ l \right] = {\mathbf{h}}_{\rm{CU},k}^H{\mathbf{x}} \left[ l \right] + z_{\rm{CU},k} \left[ l \right],
\end{equation}
where ${\mathbf{h}}_{\rm{CU},k}^H \in \mathbb{C}^{N_t \times 1}$ denotes the MISO channel vector between the BS and the $k$-th CU, and the complex-valued $z_{\rm{CU},k} \left[ l \right]$ denotes the zero-mean additive white Gaussian noise (AWGN) with the variance of each entry being ${\sigma _{\rm{CU},k}^2}$. According to the paradigm of the CI technique \cite{su2022secure, masouros2015exploiting}, the SNR per frame of the $k$-th CU is given as

\begin{equation}
    {\text{SN}}{{\text{R}}_{\rm{CU}, k}} = \frac{{{\Bbb E}\left[ {{{\left| {{\bf{h}}_{\rm{CU},k}^H{\bf{x}}\left[l\right]} \right|}^2}} \right]}}{{\sigma _{\rm{CU},k}^2}}.
\end{equation}

On the other hand, the sensing signal model can be mathematically expressed as follows:
\begin{equation}
    \mathbf{Y}_S = \mathbf{H}_S \left(\bm{\eta } \right) \mathbf{X} + \mathbf{Z}_S,
\end{equation}
where $\mathbf{Y}_S\in\mathbb{C}^{N_r\times L}$, $\mathbf{Z}_S$ represents the AWGN with zeros-mean complex-value elements each with the variance of $\sigma_S^2$, and $\mathbf{H}_S \in {\mathbb{C}^{N_r \times N_t}}$ denotes the target response matrix, which is a function of the physical parameters $\bm{\eta} \in {\mathbb{R}^M}$ to be estimated, including range, angle, and Doppler, with $M$ denoting the number of parameters to be estimated. In this paper, we consider a particular case of channel matrix $\mathbf{H}_S$, which is expressed as
\begin{equation}
    {{\mathbf{H}}_S} = \sum\limits_{n = 1}^{{K_{tar}}} {{\alpha _n}{\mathbf{b}}\left( {{\theta _n}} \right){{\mathbf{a}}^H}} \left( {{\theta _n}} \right),
\end{equation}
where $\alpha_n$ denotes the channel coefficient of each target, consisting of both the radar cross section (RCS) and path loss, which obeys the complex Gaussian distribution, and $\mathbf{a}\left( \theta \right)$, $\mathbf{b}\left( \theta \right)$ represent the transmit and receive steering vector, respectively. The received signal at the $n$-th target/Eve is accordingly written as
\begin{equation}
    {{\mathbf{y}}_{\rm{E},n}} = {\beta _n}{{\mathbf{a}}^H}\left( {{\theta _n}} \right){\mathbf{X}} + {{\mathbf{e}}_n},
\end{equation}
where $\beta_n, \forall\;n$ denotes the path loss of the $n$-th target/Eve, ${{\bf{e}}_n}$ denotes the zero-mean AWGN vector, with the variance of each entry being ${\sigma _{\rm{E},n}^2}$. 

Given the channel model (4), we define the vector with the unknown targets' parameters $\bm{\eta} = \left[ \text{Re}\left\{ \bm{\alpha} \right\}, \text{Im}\left\{ \bm{\alpha} \right\}, \bm{\theta}  \right] \in \mathbb{C}^{N\times3}$, with $ \bm{\alpha} = \left[ \alpha_1, \ldots, \alpha_N\right]^T, \bm{\theta} = \left[ \theta_1, \ldots, \theta_N\right]^T$. The steering vector and its derivative are specified as (assuming an even number of antennas) :
\begin{equation}
\begin{aligned}
        &{\mathbf{a}}\left( \theta  \right) = {\left[ {{e^{ - j\pi \frac{{{N_t} - 1}}{2}\sin \left( \theta  \right)}},{e^{ - j\pi \frac{{{N_t} - 3}}{2}\sin \left( \theta  \right)}}, \ldots ,{e^{j\pi \frac{{{N_t} - 1}}{2}\sin \left( \theta  \right)}}} \right]^T}, \hfill \\
        &{\mathbf{\dot a}}\left( \theta  \right) = {\left[ { - j\pi \frac{{{N_t} - 1}}{2}\cos \left( \theta  \right){a_1}, \ldots ,j\pi \frac{{{N_t} - 1}}{2}\cos \left( \theta  \right){a_{{N_t}}}} \right]^T},
\end{aligned}
\end{equation}
where $a_n$, with $n=1,\ldots, N_t$ denotes the $n$-th element of the steering vector $\mathbf{a}\left( \theta \right) $. Here, we choose the center of the ULA as a phase reference, such that 
\begin{equation}
    {{\mathbf{a}}^H}{\mathbf{\dot a}} = 0, {{\mathbf{b}}^H}{\mathbf{\dot b}} = 0.
\end{equation}
Accordingly, the covariance matrix of the dual-functional transmitted signal is given as 
\begin{equation}
    \mathbf{R}_x = \frac{1}{L} \mathbf{X}{\mathbf{X}^H} = \frac{1}{L}\sum\limits_{l = 1}^{{L}} { \mathbf{x}\left[ l \right]{\mathbf{x}^H}\left[ l \right]}.
\end{equation}

For the sensing performance metric, we employ the estimation mean-squared error (MSE) of $\bm{\eta}$, which is bounded by the CRB. By denoting the estimation of $\boldsymbol{\eta}$ as $\hat{\boldsymbol{\eta}}$, we have that:
\begin{equation}
    {\text{MS}}{{\text{E}}_{\bm{\eta }}}\left( {{\bm{\hat \eta }}} \right) \geq  {{\text{tr}}\left( {{\mathbf{J}}^{ - 1}} \right)},
\end{equation}
where ${{\mathbf{J}}}$ is the Bayesian Fisher Information Matrix (BFIM) of $\bm{\eta}$ which is defined as follows:
\begin{equation}
\begin{aligned}
  {{\mathbf{J}}} = &\mathbb{E}_{\bm{\eta}}\left\{ {\frac{{\partial \ln {p_{{{\mathbf{Y}}_S}|{\bm{\eta }}}}\left( {{{\mathbf{Y}}_S}|{\bm{\eta }}} \right)}}{{\partial {\bm{\eta }}}}\frac{{\partial \ln {p_{{{\mathbf{Y}}_S}|{\bm{\eta }}}}\left( {{{\mathbf{Y}}_S}|{\bm{\eta }}} \right)}}{{\partial {{\bm{\eta }}^T}}} } \right\} \hfill \\
   + &\mathbb{E}_{\bm{\eta}}\left\{ {\frac{{\partial \ln {p_{\bm{\eta }}}\left( {\bm{\eta }} \right)}}{{\partial {\bm{\eta }}}}\frac{{\partial \ln {p_{\bm{\eta }}}\left( {\bm{\eta }} \right)}}{{\partial {{\bm{\eta }}^T}}}} \right\}, \hfill \\ 
\end{aligned} 
\end{equation}
where $p_{\bm{\eta}}\left( \bm{\eta} \right)$ denotes the prior distribution of the parameters' vecror $\bm{\eta}$, and ${p_{{{\mathbf{Y}}_S}|{\bm{\eta }}}}\left( {{{\mathbf{Y}}_S}|{\bm{\eta }}} \right)$ is the probability of observing the data $\mathbf{Y}_S$ given the parameter ${\bm{\eta }}$. To derive the BFIM, we firstly let ${{\mathbf{y}}_S}{\text{ = vec}}\left( {{\mathbf{Y}}_S^T} \right)$, thus the sensing signal model can be rewritten as  
\begin{equation}
    {{\mathbf{y}}_S} = \left( {{{\mathbf{I}}_{{N_r}}} \otimes {{\mathbf{X}}^T}} \right){\text{vec}}\left( {{\mathbf{H}}_S^T} \right) + {\text{vec}}\left( {{\mathbf{Z}}_S^T} \right).
\end{equation}
Then, let ${{\mathbf{h}}_S} = {\left[ {{\text{vec}}{{\left( {{\mathbf{H}}_S^T} \right)}^T},{\text{vec}}{{\left( {{\mathbf{H}}_S^T} \right)}^H}} \right]}$ and ${\mathbf{F}} = \frac{{\partial {\mathbf{h}}_S^*}}{{\partial {\bm{\eta }}}} \in {\mathbb{C}^{K \times 2{N_t}{N_r}}}$. We further partition $\mathbf{F}$ as
\begin{equation}
    {\mathbf{F}} = \left[ {\begin{array}{*{20}{c}}
  {{{\mathbf{F}}_1},}& \ldots &{,{{\mathbf{F}}_{2{N_r}}}} 
\end{array}} \right],
\end{equation}
where ${\mathbf{F}_n} \in \mathbb{C}^{K \times N_t}$, with $n = 1, \ldots, 2N_r$. Accordingly, the BFIM can be rewritten as \cite{xiong2023fundamental}

\begin{equation}
\begin{aligned}
  &{\mathbf{J}} = \frac{L}{{\sigma _s^2}}\left\{{\mathbb{E}_{\bm{\eta }}}\left\{ {{\mathbf{F}}\left[ {\begin{array}{*{20}{c}}
  {{{\mathbf{I}}_{{N_r}}} \otimes {\mathbf{R}}_x^T}&{\mathbf{0}} \\ 
  {\mathbf{0}}&{{{\mathbf{I}}_{{N_r}}} \otimes {{\mathbf{R}}_x}} 
\end{array}} \right]{{\mathbf{F}}^H}} \right\} + {{\mathbf{J}}_P} \right\} \hfill \\
   &\;\;\;= \frac{L}{{\sigma _s^2}}\left\{{\mathbb{E}_{\bm{\eta }}}\left\{ {\sum\limits_{i = 1}^{{N_r}} {\left( {{{\mathbf{F}}_i}{\mathbf{R}}_x^T{\mathbf{F}}_i^H + {{\mathbf{F}}_{{N_r} + i}}{{\mathbf{R}}_x}{\mathbf{F}}_{{N_r} + i}^H} \right)} } \right\} + {{\mathbf{J}}_P} \right\} \hfill \\ 
\end{aligned} 
\end{equation}
where ${{\mathbf{J}}_P}$ depends on the \textit{a priori} distribution ${{p_{\mathbf{\eta }}}\left( {\mathbf{\eta }} \right)}$.

To deal with the expectation operation in (14), we define the following matrices:
\begin{subequations}
\begin{align}
    &{{\mathbf{{\bm {A}}}}_1}\left( {\mathbf{\Xi }} \right) = \sum\limits_{i = 1}^{{N_r}} {{{\mathbf{F}}_i}{\mathbf{\Xi} \mathbf{F}}_i^H}, \hfill \\
    &{{\mathbf{{\bm {A}}}}_2}\left( {\mathbf{\Xi}} \right) = \sum\limits_{i = 1}^{{N_r}} {{{\mathbf{F}}_{{N_r} + i}}{\mathbf{\Xi} \mathbf{F}}_{{N_r} + i}^H}. 
\end{align}
\end{subequations}
To derive the expectation of the later matrices, we start with (14a) and define the auxiliary matrices:
\begin{subequations}
\begin{align}
    &{{\mathbf{{\bm {B}}}}_1} = \sum\limits_{i = 1}^{{N_r}} {{\text{vec}}\left( {{{\mathbf{F}}_i}} \right){\text{vec}}{{\left( {{{\mathbf{F}}_i}} \right)}^H}},  \hfill \\
    &{{\mathbf{\bar {\bm { B}}}}_1} = \mathbb{E}\left\{ {{\mathbf{{\bm {B}}}}_1} \right\} = \sum\limits_{i = 1}^{{N_r}} {\mathbb{E}\left\{ {{\text{vec}}\left( {{{\mathbf{F}}_i}} \right){\text{vec}}{{\left( {{{\mathbf{F}}_i}} \right)}^H}} \right\}},
\end{align}
\end{subequations}
where the latter's eigenvalue decomposition is defined as:
\begin{equation}
\begin{aligned}
    {{{\mathbf{\bar {\bm {B}}}}}_1} &= {{\mathbf{U}}_1}{{\mathbf{\Lambda }}_1}{\mathbf{U}}_1^H
    = \sum\limits_{i = 1}^{{r_1}} {\left( {\sqrt {{\lambda _i}} {\mathbf{u}_i}} \right)} {\left( {\sqrt {{\lambda _i}} {\mathbf{u}_i}} \right)^H},
\end{aligned}
\end{equation}
where $\mathbf{u}_i$ denotes the corresponding eigenvector of $\lambda_i$, with $i = 1.\ldots,r_1$. We assume that $\lambda_1 \geq \lambda_2, \ldots, \lambda_{M{N_t}}$ and let $r_1$ denote the number of non-zero elements in $\mathbf{\Lambda}_1$. It can be easily shown that 
\begin{equation}
    \mathbb{E}\left\{ {{\mathbf{\bm{A}}_1}\left( \bm{\Xi}  \right)} \right\} = \sum\limits_{i = 1}^{{r_1}} {{{{\mathbf{\tilde F}}}_i}\bm{\Xi} {\mathbf{\tilde F}}_i^H},
\end{equation}
where ${{{\mathbf{\tilde F}}}_i} = \sqrt {{\lambda _i}} {\text{mat}}\left( {{\mathbf{u}_i}} \right)$.
%\begin{equation}
 %   {{{\mathbf{\tilde F}}}_i} = \sqrt {{\lambda _i}} {\text{mat}}\left( {{\mathbf{u}_i}} \right)
%\end{equation}

Likewise, we have $\mathbb{E}\left\{ {{{\mathbf{A}}_2}\left( \bm{\Xi}  \right)} \right\} = \sum\limits_{i = 1}^{{r_2}} {{{{\mathbf{\tilde G}}}_i}\bm{\Xi} {\mathbf{\tilde G}}_i^H}$, where ${{\mathbf{\tilde G}}_i} = \sqrt {{{{\mathbf{\bar \lambda }}}_i}} {\text{mat}}\left( {{{{\mathbf{\bar u}}}_i}} \right)$, as derived from (14b). To this end, the BFIM is consequently reformulated as follows:
\begin{equation}
    {{\mathbf{J}}} = \frac{L}{{\sigma _s^2}}\left( {\sum\limits_{i = 1}^{{r_1}} {{{{\mathbf{\tilde F}}}_i}{\mathbf{R}}_x^T{\mathbf{\tilde F}}_i^H}  + \sum\limits_{j = 1}^{{r_2}} {{{{\mathbf{\tilde G}}}_j}{{\mathbf{R}}_x}{\mathbf{\tilde G}}_j^H} } \right) + {{\mathbf{J}}_P}.
\end{equation}
Therefore, the BCRB with respect to $\bm{\eta}$ is accordingly given as 
\begin{equation}
    {\text{BCR}}{\text{B}} \triangleq {\text{tr}}\left( {{\mathbf{J}}^{ - 1}} \right).
\end{equation}

\section{Problem Formulation}
Given the simplified expression of the BFIM, we are now ready to formulate the optimization problem to minimize the BCRB, while conveying the received signals at CUs into the constructive region and constraining the transmit power by designing the signal matrix $\mathbf{X}$. Moreover, the received signals at targets/Eves are limited in the destructive region for the communication data security concern. Inspired by the CI-DI technique proposed in  \cite{su2022secure, masouros2015exploiting}, the BCRB minimization problem is formulation as follows
\begin{subequations}
\begin{align}
  &\mathop {\min }\limits_{\mathbf{X}}\;\; {\text{tr}}\left( {{{\mathbf{J}}^{ - 1}}} \right) \hfill \\
  &{\text{s}}{\text{.t}}{\text{.}}\;\;\frac{1}{L}\left\| {\mathbf{X}} \right\|_F^2 \leq {P_T}, \hfill \\
 & \left| {\operatorname{Im} \left( {{\mathbf{\tilde h}}_{\rm{CU},\textit{k}}^H{\mathbf{X}}} \right)} \right| \!\leq\! \left( {\operatorname{Re} \left( {{\mathbf{\tilde h}}_{\rm{CU},\textit{k}}^H{\mathbf{X}}} \right) \!-\! \sqrt {\sigma _{{\rm{CU},\textit{k}}}^2{\Gamma _{\rm{CU}, \textit{k}}}} } \right)\!\tan \phi, \forall k,  \hfill \\ 
  & \left| {{\mathop{\rm Im}\nolimits} \left( {{\beta _n}{{{\bf{\tilde a}}}^H}\left( {{\theta _n}} \right){\bf{X}}} \right)} \right|\!\ge\! \left( {{\mathop{\rm Re}\nolimits} \left( {{\beta _n}{{{\bf{\tilde a}}}^H}\left( {{\theta _n}} \right){\bf{X}}} \right) -  {\tau_{\rm{E},n}} } \right)\!\tan \phi, \forall n,
\end{align} 
\end{subequations}
where ${\mathbf{\tilde h}}_{\rm{CU},k}^H = {\mathbf{h}}_{\rm{CU},k}^Hs_k^*$, and ${{{\bf{\tilde a}}}^H}\left( {{\theta _n}} \right) = {{\bf{a}}^H}\left( {{\theta _n}} \right)s_1^*$ by taking the symbol $s_1$ as a reference. $P_T$ denotes the transmit power budget, $\Gamma_{\rm{CU},k}, \forall\;k$ is the given SNR thresholds for CUs, and ${\tau_{\rm{E},n}}$ is the given scalar for limiting the targets' received symbols in the DI region. Note that ${\tau_{\rm{E},n}}$ is generally set much smaller than the CUs' \text{SNR} threshold $\Gamma_{\rm{CU},k}, \forall\;k$.  We assume that the intended signals are $M$-Phase-shift keying (PSK) modulated, thus $\phi = \pm {\pi  \mathord{\left/ {\vphantom {\pi  M}} \right. \kern-\nulldelimiterspace} M}$. The constraint (20c) limits the signals received by CUs within the constructive region, while (20d) limits the received signals being distributed out of the constructive region. This makes correct detection more challenging for the targets by designing the received signals' constellation, meanwhile reducing the eavesdropping SINR \cite{su2022secure, xu2020rethinking, khandaker2018secure}.\footnote{The eavesdropping SINR at the $n$-th target/Eve regarding the $k$-th CU is expressed as ${\text{SINR}}_{n,k}^E = \frac{{{\Bbb E}\left[ {{{\left| {{\beta _n}{{\bf{a}}^H}\left( {{\theta _n}} \right){\bf{x}}\left[ l \right]} \right|}^2}} \right]}}{{{\Bbb E}\left[ {{{\left| {{\beta _n}{{\bf{a}}^H}\left( {{\theta _n}} \right){\bf{x}}\left[ l \right] - {s_{k,l}}} \right|}^2}} \right] + \sigma _{E,n}^2}}$, where ${{s_{k,l}}}$ denotes the desired constellation symbol for the $k$-th CU at the $l$-th time slot. It is easy to note that the ${\text{SINR}}_{n,k}^E$ is constrained once the inequality (20d) is satisfied.}

Note that the nonconvexity of problem (20) lies in the objective function and the constraint (20d). Following the method presented in \cite{su2022secure}, we divide the destructive region into three zones, that is, the inequality (20d) holds when any one of the following constraints is fulfilled. 

\noindent \textbf{case 1}: 
\begin{equation*}
{\mathop{\rm Re}\nolimits} \left( {{\beta _n}{{{\bf{\tilde a}}}^H}\left( {{\theta _n}} \right){\bf{X}}} \right) \le {\tau_{\rm{E},n}},
\end{equation*}

\noindent \textbf{case 2}: 
\begin{equation*}
    {\mathop{\rm Im}\nolimits} \left( {{\beta _n}{{{\bf{\tilde a}}}^H}\left( {{\theta _n}} \right){\bf{X}}} \right) \ge \left( {{\mathop{\rm Re}\nolimits} \left( {{\beta _n}{{{\bf{\tilde a}}}^H}\left( {{\theta _n}} \right){\bf{X}}} \right) - {\tau_{\rm{E},n}} } \right)\tan \phi 
\end{equation*}
\indent \;\;\;\;\;\;\;\;\;\;\;\;\;\;and ${\mathop{\rm Re}\nolimits} \left( {{\beta _n}{{{\bf{\tilde a}}}^H}\left( {{\theta _n}} \right){\bf{X}}} \right) > {\tau_{\rm{E},n}} $,

\noindent \textbf{case 3}:
\begin{equation*}
     - {\mathop{\rm Im}\nolimits} \left( {{\beta _n}{{{\bf{\tilde a}}}^H}\left( {{\theta _n}} \right){\bf{X}}} \right) \ge \left( {{\mathop{\rm Re}\nolimits} \left( {{\beta _n}{{{\bf{\tilde a}}}^H}\left( {{\theta _n}} \right){\bf{X}}} \right) - {\tau_{\rm{E},n}} } \right)\tan \phi 
\end{equation*}
\indent \;\;\;\;\;\;\;\;\;\;\;\;\;\;and ${\mathop{\rm Re}\nolimits} \left( {{\beta _n}{{{\bf{\tilde a}}}^H}\left( {{\theta _n}} \right){\bf{X}}} \right) > {\tau_{\rm{E},n}}. $

Till now, (20d) is rewritten as three linear constraints, that is, problem (20) is converted to three subproblems. We solve each subproblem and the one that results in the minimum value of the BCRB is the final solution to problem (20). However, the objective function is still nonconvex. In the following section, we present an efficient solver following the successive convex approximation (SCA) approach.

\section{Proposed Secure ISAC Signaling Design}
We note that the constraints in problem (20) are all convex, while the objective function is nonconvex. To this end, we define $\mathcal{Q}$ as the feasible region of problem (20), which is convex. To tackle the problem, let us denote the objective function as $f\left(\mathbf{X}\right) \triangleq {\text{tr}}\left( {{{\mathbf{J}}^{ - 1}}} \right)$. Then, we approximate the objective function by its first-order Taylor expansion near a given point $f\left({\mathbf{X'}}\right)$, yielding
\begin{equation}
    f\left( {\mathbf{X}} \right) \approx f\left( {{\mathbf{X'}}} \right) + \operatorname{Re} \left( {{\text{tr}}\left( {\nabla {f^H}\left( {{\mathbf{X'}}} \right)\left( {{\mathbf{X}} - {\mathbf{X'}}} \right)} \right)} \right),
\end{equation}
where $\nabla f\left(\cdot\right)$ denotes the gradient of $f\left(\cdot\right)$. Note that the first term in (21) is a constant, hence, we can equivalently solve the following optimization problem at the $n$-th iteration of the SCA solver:
\begin{equation}
\begin{aligned}
    &\mathop {\min }\limits_{\mathbf{X}} g\left( {\mathbf{X}} \right) \triangleq \operatorname{Re} \left( {{\text{tr}}\left( {\nabla {f^H}\left( {{{\mathbf{X}}^{n - 1}}} \right)\left( {{\mathbf{X}} - {{\mathbf{X}}^{n - 1}}} \right)} \right)} \right) \hfill \\
    &\text{s.t.}\;\; \text{(20b)\;to\;(20d)},
\end{aligned}
\end{equation}
where ${{\mathbf{X}}^{n - 1}} \in \mathcal{Q}$ is the optimal signal at the $\left(n-1\right)$-th algorithmic iteration. By solving problem (22), we obtain the optimal solution, which is denoted as $\mathbf{X}^* \in \mathcal{Q}$. Here, the term $\mathbf{X}^*-\mathbf{X}^{\left(n-1\right)}$ yields a descent direction for each iteration. By letting the variable move along the descent direction with a stepsize $\lambda$, we have
\begin{equation}
    {{\mathbf{X}}^i} = {{\mathbf{X}}^{i - 1}} + \lambda\left( {{{\mathbf{X}}^*} - {{\mathbf{X}}^{i - 1}}} \right),
\end{equation}
where the stepsize $\lambda$ may be obtained by adopting the Armijo search or the exact line search \cite{boyd2004convex}. It is noteworthy that $\mathbf{X}^i \in \mathcal{Q}$. For clarity, the SCA method applied to solving problem (20) is summarized in Algorithm 1.

\begin{algorithm}[!t]
\caption{SCA Algorithm for Solving (20)}\label{algo1}
\begin{algorithmic}[1]
\renewcommand{\algorithmicrequire}{\textbf{Input:}} 
\REQUIRE
$\mathbf{H}, P_T, \Gamma_k, \forall\;k, K_{cu}, K_{tar}, \mathbf{J}_p, \epsilon>0$, and the maximum iteration number $n_{max}$
\renewcommand{\algorithmicrequire}{\textbf{Output:}}  
\REQUIRE
$\mathbf{X}$
\renewcommand{\algorithmicrequire}{\textbf{Initialization:}}  
\REQUIRE
Initialize $\mathbf{X}^0 \in \mathcal{Q}$, and set $n=1$.
\REPEAT
\STATE
Calculate the gradient $\nabla f\left( \mathbf{X}^{n-1} \right)$.
\STATE
Rewrite the problem (20) as three subproblems by dividing the destructive region into three zones, and obtain the optimal solutions $\mathbf{X_1}^*, \mathbf{X_2}^*$, and $\mathbf{X_3}^*$.
\STATE
Update the solutions by (23), where $\lambda$ can be obtained by deploying the Armijo search or the exact line search.
\STATE
$n = n+1$.
\UNTIL $g\left( \mathbf{X}^i \right) > -\epsilon $ or $n>n_{max}$.
\STATE
Calculate the value of the objective function utilizing the obtained $\mathbf{X_1}^*, \mathbf{X_2}^*$, and  $\mathbf{X_3}^*$, and choose the one that results in the minimum BCRB as the final solution to problem (20).
\STATE
end
\end{algorithmic}
\end{algorithm}

\section{Numerical Results and Discussion}\label{sec12}
In this section, numerical results are presented based on Monte Carlo simulations of the proposed optimization technique, i.e., CI-based BCRB optimization. Without loss of generality,  we set $N_t = 12$, $N_r = 10$, and $L = 100$. The communication channel is assumed to be Rayleigh fading, where each entry of the channel gain vector $\mathbf{h}^H_{\rm{CU},k}, \forall\;k$ is subject to the standard complex Gaussian distribution. Regarding the prior distribution of the parameters to be estimated, we assume that the propagation loss $\alpha_n, \forall \; n$ in (4) obeys the complex Gaussian distribution with the variance of $\sigma^2_0$. The prior distribution of each $n$-th target's angle is assumed to be the von Mises distribution with a mean of $\mu_k$ and a standard deviation of $\sigma_{{\theta}_{k}}$, which is expressed as follows:
\begin{equation}
    f(x | \mu, \kappa) = \hfill \frac{1}{2\pi I_0(\kappa)} \exp\{\kappa \cos\left(x - \mu \right)\}, \hfill \\ 
\end{equation}
where $x$ is the circular variable (e.g., angle), $\mu$ is the mean direction (a.k.a. the location parameter), and $\kappa = \frac{1}{\sigma^2_{{\theta}_{n}}}$ is the concentration parameter, which is analogous to the inverse of the variance in a normal distribution. $I_d$ ($\kappa$) is the modified Bessel function of order $d$. Note that the FIM for Gaussian distributions is the inverse of the covariance matrix when the variables are independent. Accordingly, we have the Bayesian \textit{a priori} FIM as follows \cite{vempaty2014quantizer}
\begin{equation}
    \mathbf{J}_P =
\begin{bmatrix}
\frac{1}{2\sigma_0^2} & 0 & 0 \\
0 & \frac{1}{2\sigma_0^2} & 0 \\
0 & 0 & \kappa
\end{bmatrix}.
\end{equation}

The spatial distribution of the received signals at CUs (denoted by blue dots) and at targets/Eves (denoted by red dots) is shown in Figure 1, where QPSK and 8PSK modulated signals are taken as examples. It illustrates that the signals received by the communication users are conveyed into the constructive region by applying the CI technique, while the signals at the targets/Eves are out of the constructive region, which verifies the CI-DI technique effectively prevents the targets/Eves from receiving the right constellation of the communication data. 
\begin{figure}
  \centering
  \subfigure[QPSK]{
  \includegraphics[width=0.45\columnwidth]{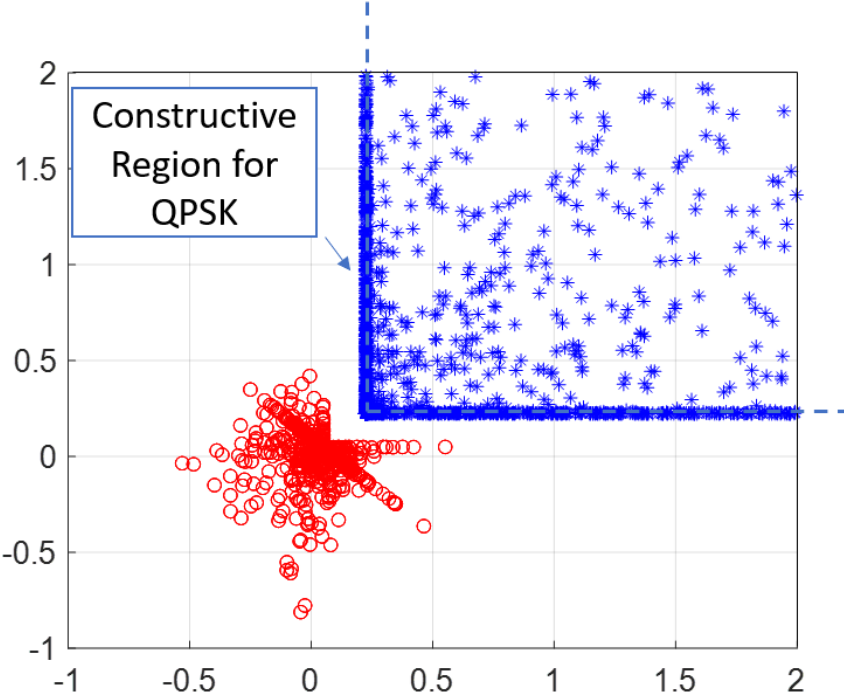}}
  \subfigure[8PSK]{
  \includegraphics[width=0.45\columnwidth]{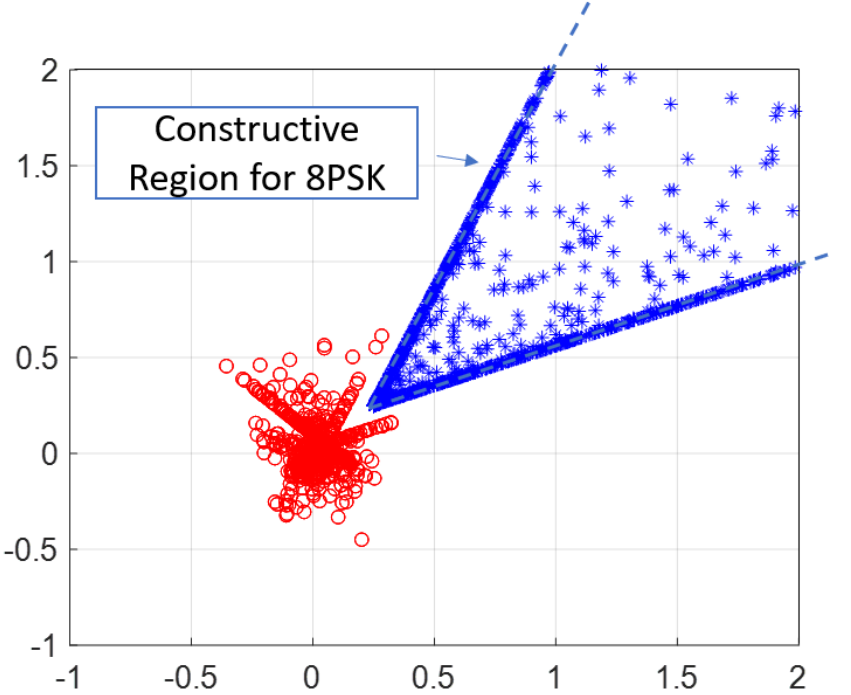}}
  \captionsetup{font={footnotesize}}
  \caption{The constellation of received signals at CUs, a) QPSK, b) 8PSK, $ K_{cu}=3, K_{tar}=2, P_0=30\;\text{dBm}, \Gamma_{\rm{CU}, k} = 15\;\text{dB}, \forall\;k, \tau_{\rm{E},n} = -5\;\text{dB}$, and $\sigma_{\theta_n} = 5^\circ, \forall\;n$.}
  \label{fig.1}
\end{figure}

In Figure 2, we demonstrate the generated beampatterns with different standard deviations of the \textit{a priori} information of the target angles. We assume that there exists in the field of interest $K_{tar} = 2$ targets located at $\theta_1 = -50^\circ$ and $\theta_2 = -20^\circ$, and the angle standard deviation is given as $1^\circ$ and $5^\circ$ in Figure 2 (a) and Figure 2 (b), respectively. Figure 2 illustrates that the main lobes pointing to targets of interest get narrow and with higher beam gain when $\sigma_{\theta_n}$ gets smaller, which implies a higher accuracy of the target angle estimation.

\begin{figure}
  \centering
  \subfigure[$\sigma_{\theta_n} = 0.5^\circ, \forall\;n$]{
  \begin{minipage}[b]{0.45\textwidth}
  \includegraphics[width=1\textwidth]{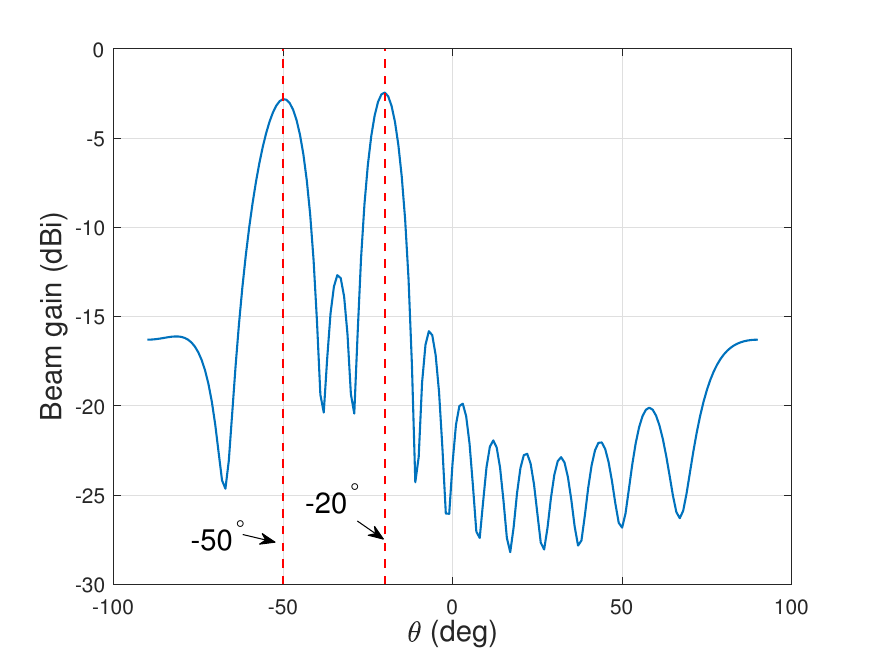}
  \end{minipage}
  }
  \subfigure[$\sigma_{\theta_n} = 5^\circ, \forall\;n$]{
    \begin{minipage}[b]{0.45\textwidth}
    \includegraphics[width=1\textwidth]{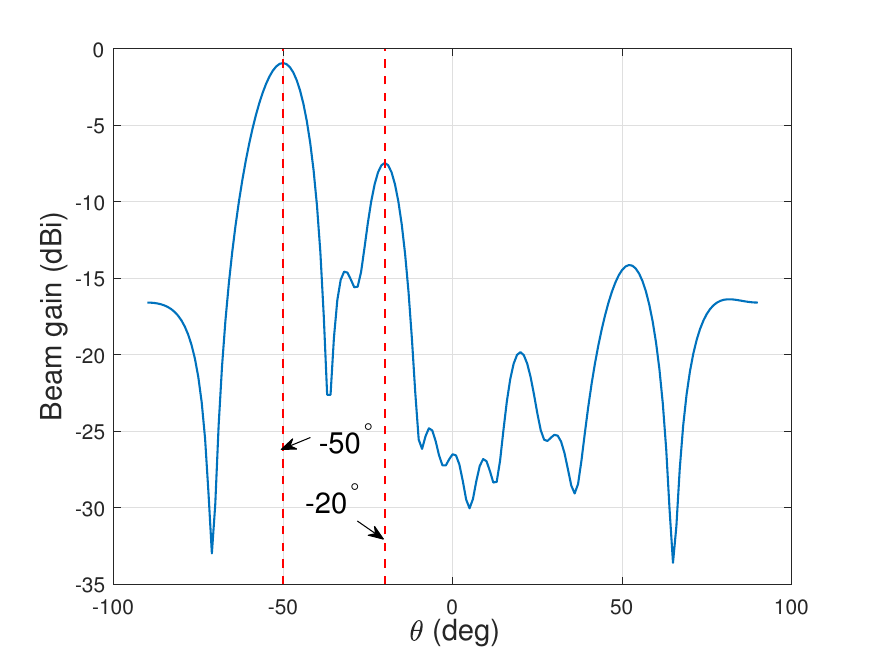}
    \end{minipage}
    }
    \captionsetup{font={footnotesize}}
    \caption{The resultant beampatterns via the proposed secure ISAC signaling design with different \textit{a priori} information of the angle, a) standard deviation is $1^\circ$, b) standard deviation is $5^\circ$, $K_{cu}=3, K_{tar}=2, P_0=30\;\text{dBm}$, $\Gamma_{\rm{CU},k} = 20\;\text{dB} \forall\;k$.} \label{fig.2}
\end{figure}
\begin{figure}
  \centering
  \includegraphics[width=0.55\textwidth]{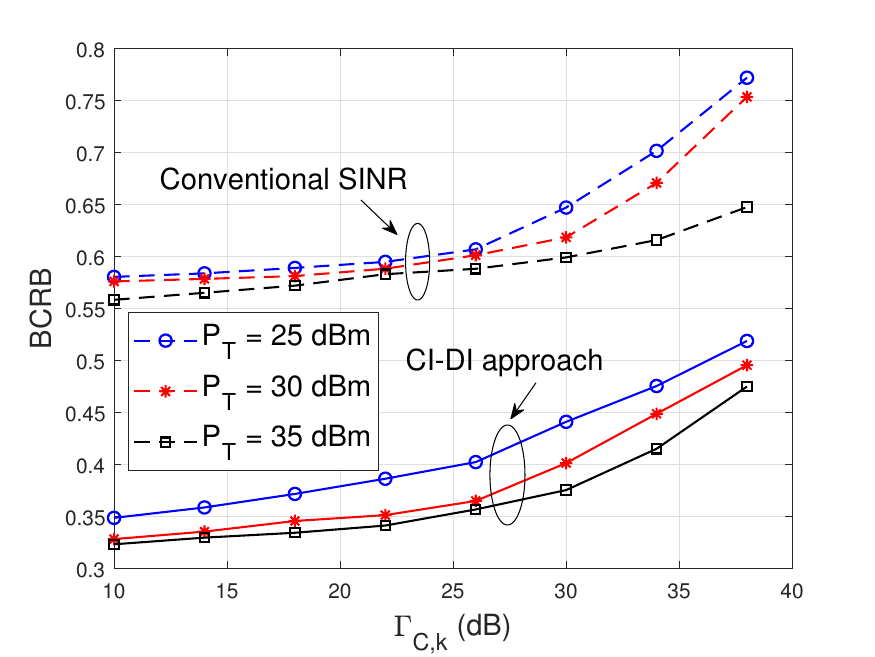}
  \captionsetup{font={footnotesize}}
  \caption{Performance tradeoff between the communication and sensing systems with different power budgets, for $K_{cu}=3 $ and $ K_{tar}=2$.} 
  \label{fig.3}
\end{figure}
Furthermore, Figure 3 shows the tradeoff between the communication and the sensing performances. It is obvious that with the improvement of the communication QoS, the CRB increases. In the block-level precoding, we leverage the conventional block-level SINR of the $k$-th CU as 
\begin{equation}
    {\rm{SIN}}{{\rm{R}}_k} = \frac{{{{\left| {{\bf{h}}_k^H{{\bf{w}}_k}} \right|}^2}}}{{\sum\nolimits_{i = 1,i \ne k}^{{K_{cu}}} {{{\left| {{\bf{h}}_k^H{{\bf{w}}_i}} \right|}^2}}  + \sigma _{\rm{CU},k}^2}},
\end{equation}
where the signal matrix $\mathbf{X}$ in (3) can be written as $\mathbf{X} = \mathbf{WS}$, with $\mathbf{W}$ denoting the precoding matrix and $\mathbf{w}_k$ denoting the $k$-th entry of the precoding matrix $\mathbf{W}$ corresponding to the $k$-th CU, and the transmit signal vector $\mathbf{s}$ is a set to include QPSK-modulated symbols. The block-level precoding scheme is set as a benchmark in Figure 3 and the numerical results are generated by replacing (20c) with the constraint ${\rm{SIN}}{{\rm{R}}_k} \ge { \Gamma _k}$. It indicates that the proposed CI-DI-based design outperforms the block-level precoding technique, due to the reason that the block-level design consumes more power to reach the same SINR/SNR threshold, i.e., $\Gamma_k$.

Figure 4 depicts the average SER at CUs and targets/Eves versus the SNR threshold $\Gamma_{\rm{CU},k}$. By imposing the DI constraints, we note that the SER at targets/Eves is close to one, which indicates that the communication data is effectively protected from being decoded by the targets/Eves. Besides, it also illustrated that the SER at CUs gets lower with an increasing power budget.
\begin{figure}
  \centering
  \includegraphics[width=0.55\textwidth]{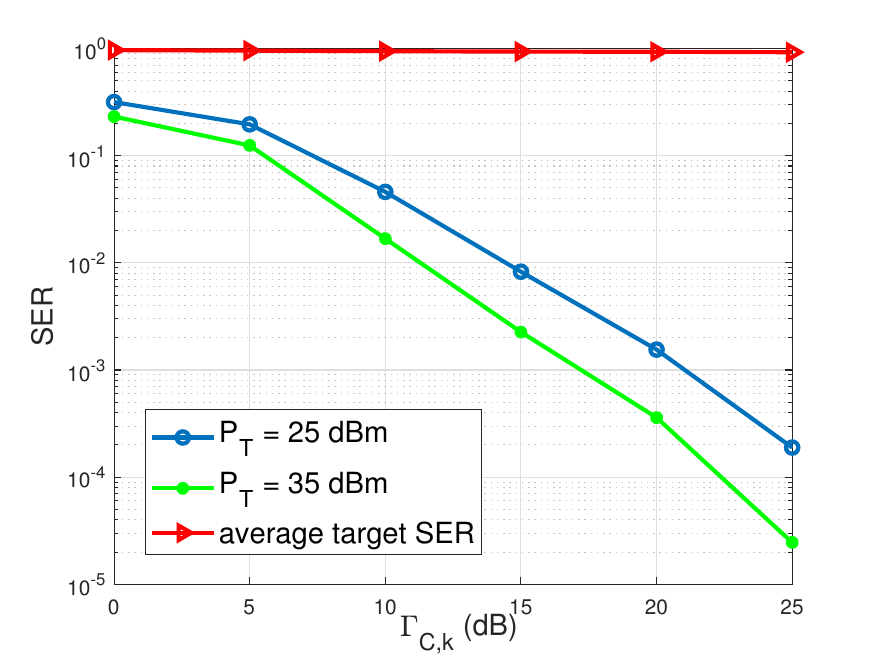}
  \captionsetup{font={footnotesize}}
  \caption{SER of CUs and average SER of targets/Eves versus the SNR threshold $\Gamma_{\rm{CU},k}$ with different power budgets for $K_{cu}=3$  and $K_{tar}=2$.} 
  \label{fig.4}
\end{figure}

\section{Conclusion}\label{sec13}

In this paper, we presented a novel symbol-level signaling design algorithm for ISAC systems aiming at ensuring communication data security. The proposed design exploits the CI-DI technique, while sensing performance was measured by the BCRB and its PLS capability was quantified by the SER. Our optimization problem formulation deals with the BCRB minimization, while conveying the received signals at CUs into the constructive region and making sure the received signals at targets/Eves fall into the destructive region. The presented numerical results verified that the CI-DI technique effectively protects communication data security. It was also showcased that the proposed symbol-level precoding technique outperforms the block-level precoding design. It was also demonstrated that the resultant beampattern with the proposed design yields improved sensing performance (i.e., narrower main beam with higher beam gain) when a priori statistical information of the unknown targets' parameters is known accurately. 

\section*{Acknowledgments}
This work was supported in part by the Engineering and Physical Sciences Research Council (EPSRC) under Grant EP/S028455/1, in part by the National Natural Science Foundation of China under Grant 62101234, Grant U20B2039, Grant 61831008 and Grant 62027802, in part by the Young Elite Scientist Sponsorship Program by CAST under Grant No. YESS20210055, in part by the China Scholarship Council (CSC), and in part by Smart Networks and Services Joint Undertaking (SNS JU) project 6G-DISAC under the European Union's Horizon Europe research and innovation program under Grant Agreement no. 101139130.

%\newpage

%\section{Biography Section}
%If you have an EPS/PDF photo (graphicx package needed), extra braces are
 %needed around the contents of the optional argument to biography to prevent
 %the LaTeX parser from getting confused when it sees the complicated
 %$\backslash${\tt{includegraphics}} command within an optional argument. (You can create
 %your own custom macro containing the $\backslash${\tt{includegraphics}} command to make things
 %simpler here.)
 
%\vspace{11pt}

%\bf{If you include a photo:}\vspace{-33pt}
%\begin{IEEEbiography}[{\includegraphics[width=1in,height=1.25in,clip,keepaspectratio]{fig1}}]{Michael Shell}
%Use $\backslash${\tt{begin\{IEEEbiography\}}} and then for the 1st argument use $\backslash${\tt{includegraphics}} to declare and link the author photo.
%Use the author name as the 3rd argument followed by the biography text.
%\end{IEEEbiography}

%\vspace{11pt}

%\bf{If you will not include a photo:}\vspace{-33pt}
%\begin{IEEEbiographynophoto}{John Doe}
%Use $\backslash${\tt{begin\{IEEEbiographynophoto\}}} and the author name as the argument followed by the biography text.
%\end{IEEEbiographynophoto}

\ifCLASSOPTIONcaptionsoff
  \newpage
\fi

% trigger a \newpage just before the given reference
% number - used to balance the columns on the last page
% adjust value as needed - may need to be readjusted if
% the document is modified later
%\IEEEtriggeratref{8}
% The "triggered" command can be changed if desired:
%\IEEEtriggercmd{\enlargethispage{-5in}}

% references section

% can use a bibliography generated by BibTeX as a .bbl file
% BibTeX documentation can be easily obtained at:
% http://mirror.ctan.org/biblio/bibtex/contrib/doc/
% The IEEEtran BibTeX style support page is at:
% http://www.michaelshell.org/tex/ieeetran/bibtex/
\bibliographystyle{IEEEtran}
% argument is your BibTeX string definitions and bibliography database(s)
\bibliography{IEEEabrv,CEP_REF}

\end{document}